# Dufour and Soret effects on MHD flow of Williamson fluid over an infinite rotating disk with anisotropic slip

*Najeeb Alam Khan[1], Faqiha Sultan[2]*

[1] *Department of Mathematics, University of Karachi, Karachi 75270, Pakistan*

[2]*Department of Sciences and Humanities, National University of Computer and Emerging Sciences, Karachi 75030, Pakistan*

## Abstract

This study deals with the investigation of MHD flow of Williamson fluid over an infinite rotating disk with the effects of Soret, Dufour, and anisotropic slip. The anisotropic slip and the Soret and Dufour effects are the primary features of this study, which greatly influence the flow, heat and mass transport properties. In simultaneous appearance of heat and mass transfer in a moving fluid, the mass flux generated by temperature gradients is known as the thermal-diffusion or Soret effect and the energy flux created by a composition gradient is called the diffusion-thermo or Dufour effect, however, difference in slip lengths in streamwise and spanwise directions is named as anisotropic slip. The system of nonlinear partial differential equations (PDEs), which governs the flow, heat and mass transfer characteristics, is transformed into ordinary differential equations (ODEs) with the help of von Kármán similarity transformation. A numerical solution of the complicated ODEs is carried out by a MATLAB routine bvp4c. The obtained results of velocity, temperature, concentration, and some physical quantities are displayed through graphs and tables, and their physical discussion is presented.

## 1. Introduction

In recent years, the analysis of non-Newtonian fluids has gripped the major consideration because of numerous applications in industry and engineering. The most common example of the fluids encountered in the study of non-Newtonian fluids are pseudoplastic fluids, which appear in the preparation of emulsion sheets such as photographic films, polymer sheets extrusion, flow of plasma and blood, etc. The rheological properties of such fluids cannot be explained by Navier Stokes equations alone, therefore, to overcome this deficiency, several rheological models have been presented such as the Carreau model, power law model, Ellis model, and Cross model. In 1929, Williamson [1] presented a detailed discussion on pseudoplastic materials, modeled a constitutive equation that defines the flow characteristics of pseudoplastic fluids, and experimentally validated the results. The resemblance of this model of non-Newtonian fluid to the blood flow has captivated the attention of researchers as it essentially exhibits the behavior of blood flow. Many valuable works have constantly been added to this field in recent years. Nadeem, et al. [2] presented a study on the two dimensional flow of a Williamson fluid over a stretching sheet. Khan, et al. [3] investigated the effects of chemical reaction on the boundary layer flow of Williamson fluid. Zehra, et al. [4] proposed numerical solutions of the Williamson fluid flow over an inclined channel with pressure dependent viscosity. Malik, et al. [5] presented Keller box solution of the homogeneous-heterogeneous reactions on Williamson fluid over a stretching cylinder. Recently, Malik, et al. [6] have numerically investigated the effects of variable thermal conductivity and heat generation/absorption on Williamson fluid flow and heat transfer.

The exact solution of the momentum equations governing the steady flow driven by an infinitely rotating disk, which rotates with uniform angular velocity was firstly proposed by Kármán [7]. The flow is described by the absence of pressure gradient in radial direction close to the disk that balance out the centrifugal forces so the fluid spirals outward. The disk works as a centrifugal fan, the flow generating from the disk get supplanted by an axial flow bounced back to the surface of the disk. The study of Newtonian fluids over a rotating disk problems have gained massive attention from numerous authors over many decades. The non-Newtonian flow over a rotating disk is a significant problem and possess many applications in engineering, but much less attention has been directed towards the investigation of non-Newtonian rotating disk problem. Some of the recent studies include: the effect of magnetic field on laminar flow of non-Newtonian Eyring-Powell fluid over a rotating disk, investigated by Khan, et al. [8], the stability of the boundary layer for the flow of non-Newtonian fluids due to a rotating disk, proposed by Griffiths, et al. [9], and the effects of double diffusion on the unsteady MHD flow of couple stress fluid and heat transfer over a rotating disk, presented by Khan, et al. [10].

Latest development in micro- and nanotechnology has exposed the possibility of using micro- and nanodevices in a wide range of applications. In micro- and nanofluidic devices, surface properties play a major role. The hydrophobic or hydrophilic nature of surfaces conjured by using micro- or nanotechnology has attracted much attention lately. It has been appeared that on hydrophobic surfaces, the slip velocity results in significant drag reduction in microchannel flows. Recently, superhydrophobic surfaces have captivated much attention as they have the possibility of accomplishing a substantial skin-friction drag reduction in turbulent flows. Several engineering applications of these surfaces have been explored lately, such as microfluidic devices, self-cleaning, skin-friction drag reduction, anti-biofouling, and anti-fogging [11-13]. The flow over some surfaces shows evident slip, denying the traditional no-slip condition and in recent years, some development have been made in studying such flows. Ng and Wang [11] investigated the implications of superhydrophobic surfaces and studied the Stokes shear flow over a grating. Later, for Stokes flow they obtained an effective slip over a surface with two-or three-dimensional projected patterns by using a semi-analytical technique [14]. Busse and Sandham [15] investigated the influence of anisotropic Naiver slip-length boundary condition on turbulent channel flow. Wang [16] studied an axisymmetric stagnation flow over a moving plate with different streamwise and spanwise slip coefficients. Recently, Cooper, et al. [17] have presented a theoretical study on the effect of anisotropic and isotropic slip on the stability of the boundary layer flow over a rotating disk.

Many technical and industrial applications are based on the forced and free convection, for example, cooling of electronic devices cooled by fans, cooling of nuclear reactors during emergency shutdown, and heat exchangers low-velocity environments. Moreover, cross-diffusion is caused by the simultaneous occurrence of heat and mass transfers that affect each other. The mass transfer affected by the temperature gradient is known as the Soret effect, whereas the heat transfer affected by the concentration gradient is known as the Dufour effect. Osalusi, et al. [18] investigated the Dufour and Soret effects on steady convective flow due to a rotating disk with heat and mass transfer in the presence of magnetic field, slip condition, Ohmic heating, and viscous dissipation. Shateyi and Motsa [19] studied the boundary layer flow over an unsteady stretching surface with Hall current and Soret-Dufour effects. Narayana and Sibanda [20] the double diffusive effects on the flow due to a cone. Recently, Khan and Sultan [21] have presented the flow of Eyring-Powell fluid over a cone bounded by a porous medium with Soret and Dufour effects.

By surveying the literature, it is observed that the flow over a rotating disk, Soret and Dufour effects, and the effect of anisotropic slip on the flow, heat and mass transfer characteristics in Williamson fluid have not been analyzed so far, although, these flow characteristics have several applications in oceanography, rotating machinery, computer storage devices, skin-friction drag reduction, nanotechnology, electronic and nuclear devices, etc. In this work, an effort has been made to overcome this deficiency and investigate the influence of anisotropic slip on the flow, heat and mass transfer in MHD flow over a rotating disk with Soret and Dufour effects. The von Kármán similarity transformation has been used to transform the ODEs governing the momentum, heat and mass transfer characteristics into PDEs. Considering all the above effects in unison, the model become mathematically more complicated that it could not be solved analytically. A numerical solution of this model is obtained by using a MATLAB routine bvp4c.

## 2. Mathematical Model

The fluid of interest is pseudoplastic fluid, which follows the Williamson rheological model. The extra stress tensor in a Williamson fluid is given as:

$$\tau = \left(\mu_\infty + \left(\mu_0 - \mu_\infty\right)\left(1 - \Gamma\dot{\gamma}\right)^{-1}\right)A \tag{1}$$

where $\mu_0$ and $\mu_\infty$ are the zero and infinite shear rates viscosity, respectively, $\Gamma$ is the time constant, $A = \nabla V + \left(\nabla V\right)^t$ is the rate-of-strain-tensor, $\nabla$ is the differential operator, and $V$ is the velocity $\left(u, v, w\right)$ vector defined in the cylindrical coordinates $\left(r, \theta, z\right)$, and $\dot{\gamma}$ is the second invariant of rate-of-strain-tensor that can be defined as:

$$\dot{\gamma} = \sqrt{\frac{1}{2} tr A^2} \tag{2}$$

Using the first order Taylor's series approximation of $\left(1 - \Gamma\dot{\gamma}\right)^{-1} \cong \left(1 + \Gamma\dot{\gamma}\right)$, with $\Gamma^2 <<< 1$, Eq. (1) takes the form:

$$\tau = \left(\mu_0 + \left(\mu_0 - \mu_\infty\right)\Gamma\dot{\gamma}\right)A \tag{3}$$

The shear components of the Williamson fluid in cylindrical polar coordinates become:

$$\tau_{rr} = 2\left(\mu_0 + (\mu_0 - \mu_\infty)\Gamma\dot{\gamma}\right)\left(\frac{\partial u}{\partial r}\right), \quad \tau_{r\theta} = \tau_{\theta r} = \left(\mu_0 + (\mu_0 - \mu_\infty)\Gamma\dot{\gamma}\right)\left(\frac{1}{r}\frac{\partial u}{\partial \theta} + \frac{\partial v}{\partial r} - \frac{v}{r}\right),$$

$$\tau_{\theta\theta} = 2\left(\mu_0 + (\mu_0 - \mu_\infty)\Gamma\dot{\gamma}\right)\left(\frac{1}{r}\frac{\partial v}{\partial \theta} + \frac{u}{r}\right), \quad \tau_{rz} = \tau_{zr} = \left(\mu_0 + (\mu_0 - \mu_\infty)\Gamma\dot{\gamma}\right)\left(\frac{\partial u}{\partial z} + \frac{\partial w}{\partial r}\right), \quad (4)$$

$$\tau_{zz} = 2\left(\mu_0 + (\mu_0 - \mu_\infty)\Gamma\dot{\gamma}\right)\left(\frac{\partial w}{\partial z}\right), \qquad \tau_{\theta z} = \tau_{z\theta} = \left(\mu_0 + (\mu_0 - \mu_\infty)\Gamma\dot{\gamma}\right)\left(\frac{1}{r}\frac{\partial w}{\partial \theta} + \frac{\partial v}{\partial z}\right).$$

where,

$$\dot{\gamma} = \sqrt{2\left(\frac{\partial u}{\partial r}\right)^2 + \left(\frac{\partial v}{\partial r} + \frac{1}{r}\frac{\partial u}{\partial \theta} - \frac{v}{r}\right)^2 + \left(\frac{\partial u}{\partial z} + \frac{\partial w}{\partial r}\right)^2 + 2\left(\frac{1}{r}\frac{\partial v}{\partial \theta} + \frac{u}{r}\right)^2 + \left(\frac{1}{r}\frac{\partial w}{\partial \theta} + \frac{\partial v}{\partial z}\right)^2 + 2\left(\frac{\partial w}{\partial z}\right)^2} \quad (5)$$

The equations of continuity, momentum, temperature, and concentration are defined as:

$$\nabla \cdot V = 0 \qquad\qquad\qquad\qquad\qquad\qquad\qquad\qquad\qquad\qquad\qquad (6)$$

$$\rho(V \cdot \nabla)V = -\nabla p + \nabla \cdot \tau \qquad\qquad\qquad\qquad\qquad\qquad\qquad (7)$$

$$\rho C_p (V \cdot \nabla)T = k \nabla^2 T \qquad\qquad\qquad\qquad\qquad\qquad\qquad (8)$$

$$(V \cdot \nabla)C = D \nabla^2 C \qquad\qquad\qquad\qquad\qquad\qquad\qquad\qquad (9)$$

In above equations, $\rho$ is the fluid density, $p$ is the hydrostatic pressure, $C_P$ is the specific heat at constant pressure, $T$ and $C$ are the species temperature and concentration, $k$ is the thermal conductivity, and $D$ is the molecular diffusivity.

### 2.1    Problem Formulation

Consider a steady, incompressible, three dimensional axisymmetric flow of Williamson fluid over a rotating disk with heat and mass transfer. The flow is produced by a disk with anisotropic slip rotating infinitely with constant angular velocity $\Omega$, extended infinitely in the positive half-space of the disk $(z > 0)$ and initially placed at $z = 0$. A uniform external magnetic field of strength $B_0$ is applied normal to the disk. The schematic diagram of the problem is presented in Fig. 1. Both the temperature $T_w$ and concentration $C_w$ are kept constant at the surface of the rotating disk. The fluid has a uniform ambient temperature $T_\infty$ and ambient concentration $C_\infty$ at a constant pressure $P_\infty$.

Under the assumption of long wavelength, low Reynolds number approximation, and along with the boundary conditions, the system of equations that govern the flow, heat and mass transfer from Eqs. (6)-(9) is obtained as:

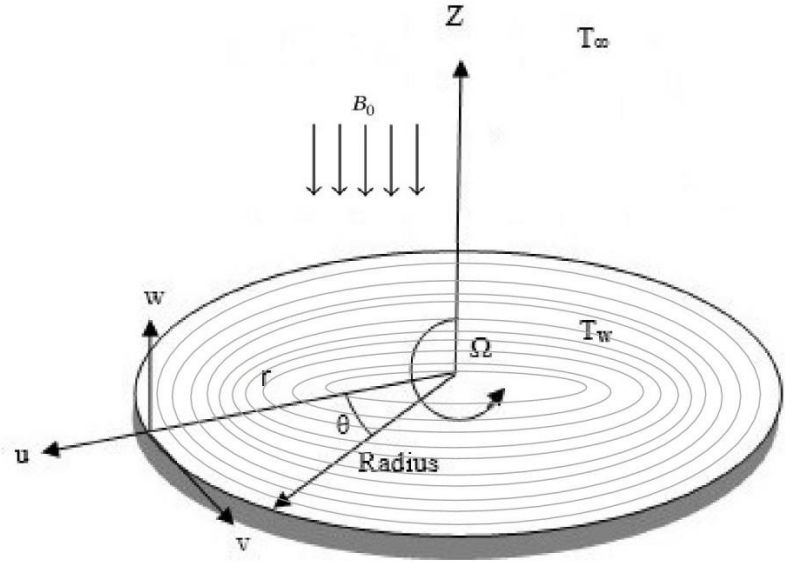

Fig. 1: Physical model of the fluid flow and coordinate system

$$\frac{\partial u}{\partial r} + \frac{u}{r} + \frac{\partial w}{\partial z} = 0 \tag{10}$$

$$\rho \left( u \frac{\partial u}{\partial r} - \frac{v^2}{r} + w \frac{\partial u}{\partial z} \right) = -\frac{\partial p}{\partial r} + \frac{\partial \tau_{rr}}{\partial r} + \frac{\partial \tau_{zr}}{\partial z} + \frac{\tau_{rr} - \tau_{\theta\theta}}{r} - \sigma B_0^2 u \tag{11}$$

$$\rho\left(u\,\frac{\partial v}{\partial r}+\frac{u\,v}{r}+w\,\frac{\partial v}{\partial z}\right)=\frac{\partial\,\tau_{r\theta}}{\partial\,r}+\frac{\partial\,\tau_{z\theta}}{\partial\,z}+2\,\frac{\tau_{r\theta}}{r}-\sigma\,B_0^2 v \tag{12}$$

$$\rho\left(u\,\frac{\partial w}{\partial r}+w\,\frac{\partial w}{\partial z}\right)=-\frac{\partial\,p}{\partial\,z}+\frac{\partial\,\tau_{rz}}{\partial\,r}+\frac{\partial\,\tau_{zz}}{\partial\,z}+\frac{\tau_{rz}}{r} \tag{13}$$

$$u\,\frac{\partial T}{\partial r}+w\,\frac{\partial T}{\partial z}=\frac{k}{\rho\,C_P}\left(\frac{\partial^2 T}{\partial r^2}+\frac{1}{r}\frac{\partial T}{\partial r}+\frac{\partial^2 T}{\partial z^2}\right)+\frac{D_m K_T}{C_s C_P}\left(\frac{\partial^2 C}{\partial r^2}+\frac{1}{r}\frac{\partial C}{\partial r}+\frac{\partial^2 C}{\partial z^2}\right) \tag{14}$$

$$u\,\frac{\partial C}{\partial r}+w\,\frac{\partial C}{\partial z}=D\left(\frac{\partial^2 C}{\partial r^2}+\frac{1}{r}\frac{\partial C}{\partial r}+\frac{\partial^2 C}{\partial z^2}\right)+\frac{D_m K_T}{T_m}\left(\frac{\partial^2 T}{\partial r^2}+\frac{1}{r}\frac{\partial T}{\partial r}+\frac{\partial^2 T}{\partial z^2}\right) \tag{15}$$

The boundary conditions associated to the model are:

$$u=k_1\,\tau_{rz},\quad v=r\Omega+k_2\,\tau_{\theta z},\quad w=0,\quad T=T_w,\quad C=C_w\ \text{ at }\ z=0,$$

$$u\to 0,\quad v\to 0,\quad w\to 0,\quad T\to T_\infty,\quad C\to C_\infty,\quad P\to 0\ \text{ as }\ z\to\infty. \tag{16}$$

Where $K_T$ is the thermal-diffusion ratio, $C_s$ is the susceptibility of concentration, $D_m$ is the effective diffusivity rate of mass, $T_m$ is the mean temperature of the fluid, and $k_1$ and $k_2$ are the radial and tangential slip coefficients, respectively.

### 2.2    Similarity Transformation

To obtain the dimensionless continuity, momentum, energy, and concentration equations, the similarity solution of Navier-Stokes flow obtained by Kármán [7] is used that can be defined as:

$$f(\eta) = \frac{u}{r\,\Omega}\,,\ \ g(\eta) = \frac{v}{r\,\Omega}\,,\ \ h(\eta) = \frac{w}{\sqrt{\nu\,\Omega}}\,,\ \ P(\eta) = \frac{p}{\rho\,\nu\,\Omega}\,,\ \ \Theta(\eta) = \frac{C - C_\infty}{C_w - C_\infty}\,,$$

$$\phi(\eta) = \frac{C - C_\infty}{C_w - C_\infty}\,,\quad \eta = \sqrt{\frac{\Omega}{\nu}}\,z\ \ \text{and}\ \ \mathrm{Re} = \frac{\Omega\,r^2}{\nu} \tag{17}$$

where $f$, $g$, $h$ are the dimensionless radial, axial, and tangential functions of dimensionless velocity, $\eta$ is the dimensionless distance from the surface of the disk, $\Theta$ is the dimensionless temperature, $\phi$ is the dimensionless concentration, $P$ is the dimensionless dynamic pressure in the fluid above the disk, $\nu = \dfrac{\mu_0}{\rho}$ is the kinematic viscosity, and $\mathrm{Re}$ is the Reynolds number. Using the above similarity transformation, Eqs. (10)-(15) are now transformed as:

$$2f + h' = 0, \tag{18}$$

$$\left(1 + We\dot{\gamma}\right)f'' - f^2 + g^2 - h\,f' - Mf + \frac{We\,\mathrm{Re}}{\dot{\gamma}^3}\left(f''f'^2 + f'g'g''\right) + \frac{2We}{\dot{\gamma}^3}\left(3ff'^2 + fg'^2 + f'h'h''\right) = 0 \tag{19}$$

$$\left(1 + We\dot{\gamma}\right)g'' - h\,g' - 2f\,g - Mg + \frac{We\,\mathrm{Re}}{\dot{\gamma}^3}\left(f'g'f'' + g'^2g''\right) + \frac{2We}{\dot{\gamma}^3}\left(2f\,f'g' + g'h'h''\right) = 0 \tag{20}$$

$$P' + h\,h' - 2\left(1 + We\dot{\gamma}\right)\left(f' + h''\right) - \frac{We\,\mathrm{Re}}{\dot{\gamma}^3}\left(f'^3 + f'g'^2 + f'h'f'' + h'g'g''\right) - \frac{4We}{\dot{\gamma}^3}\left(2f\,f'h' + h'^2h''\right) = 0 \tag{21}$$

$$\Theta'' - \mathrm{Pr}\,h\,\Theta' + D_f\,\phi'' = 0 \tag{22}$$

$$\phi'' - Sch\,\phi' + S_r\Theta'' = 0 \tag{23}$$

where,

$$\dot{\gamma} = \sqrt{4f^2 + \mathrm{Re}\left(f'^2 + g'^2\right) + 2h'^2} \tag{24}$$

In above equations, $We = \dfrac{\Omega \Gamma (\mu_0 - \mu_\infty)}{\mu_0}$ is the Weissenberg number, $M = \dfrac{\sigma B_0^2}{\rho \Omega}$, is the magnetic field parameter, $Pr = \dfrac{\mu_0 C_P}{k}$ is the Prandtl number, $D_f = \dfrac{\rho D K_T}{k C_s} \dfrac{(C_w - C_\infty)}{(T_w - T_\infty)}$ is the Dufour parameter, and $S_r = \dfrac{D_m K_T}{D T_m} \dfrac{T_w - T_\infty}{C_w - C_\infty}$ is the Soret parameter.

The boundary conditions defined in Eq. (16) are transformed as:

$$f(0) = \lambda_1 \, f'(0) \left( 1 + We \sqrt{4(f(0))^2 + \mathrm{Re}\left((f'(0))^2 + (g'(0))^2\right) + 2(h'(0))^2} \right)$$

$$g(0) = 1 + \lambda_2 \, g'(0) \left( 1 + We \sqrt{4(f(0))^2 + \mathrm{Re}\left((f'(0))^2 + (g'(0))^2\right) + 2(h'(0))^2} \right)$$

$$h(0) = 0, \;\; P(0) = 0 \quad \Theta(0) = 1, \quad \phi(0) = 1,$$

$$f(\infty) = 0, \;\; g(\infty) = 0, \quad \Theta(\infty) = 0, \quad \phi(\infty) = 0.$$

(25)

Where $\lambda_1 = k_1 \, \mu_0 \sqrt{\dfrac{\Omega}{\nu}}$ is the radial slip parameter and $\lambda_2 = k_2 \, \mu_0 \sqrt{\dfrac{\Omega}{\nu}}$ is the tangential slip parameter.

### 2.3    Physical Quantities

The shear stress coefficients, moment of friction, the local Nusselt number $N_u$, and the local Sherwood number $S_h$ are some important physical quantities in von Kármán flows with heat and mass transfer. The radial and tangential shear stress coefficients at the surface of the disk for Williamson fluid are:

$$\left( C_f, \; C_g \right) = \left( \dfrac{\tau_{rz}}{\mu_0 \Omega}, \; \dfrac{\tau_{\theta z}}{\mu_0 \Omega} \right) \Bigg|_{z=0}$$

(26)

Thus, the local skin-friction coefficients are obtained as:

$$\frac{C_f}{\sqrt{\text{Re}}} = \left(1 + We\sqrt{4(f(0))^2 + \text{Re}\left((f'(0))^2 + (g'(0))^2\right) + 2(h'(0))^2}\right)f'(0) \tag{27}$$

and

$$\frac{C_g}{\sqrt{\text{Re}}} = \left(1 + We\sqrt{4(f(0))^2 + \text{Re}\left((f'(0))^2 + (g'(0))^2\right) + 2(h'(0))^2}\right)g'(0) \tag{28}$$

Another interesting quantity in flows over rotating disk is the turning moment for the disk with fluid on both sides, known as the moment of friction. It is obtained from the tangential velocity profile by integrating the shear stresses over the disk surface. The moment coefficient $C_m$ for Williamson fluid over a rotating disk is obtained as:

$$C_m = \frac{-2\pi}{\sqrt{\text{Re}}} \left(1 + We\sqrt{4f^2 + \text{Re}\left(f'^2 + g'^2\right) + 2h'^2}\right)g' \tag{29}$$

The heat and mass fluxes at the surface of the rotating disk are given by:

$$\left(q_w, M_w\right) = \left(-k\left(\frac{\partial T}{\partial z}\right), -D\left(\frac{\partial C}{\partial z}\right)\right)\Bigg|_{z=0} \tag{30}$$

Thus, the Nusselt and Sherwood numbers are obtained as:

$$\left(N_u, S_h\right) = -\sqrt{\text{Re}}\left(\Theta'(0), \phi'(0)\right) \tag{31}$$

## 3. Numerical Method

The momentum, heat and mass transfer properties of Williamson fluid over a rotating disk with Soret and Dufour effects have not been studied yet because of the complexity of the system of governing equations. The governing system is highly nonlinear consisting of coupled ordinary differential equations that are quite difficult to solve analytically.

Moreover, the presence of slip terms in the boundary conditions make the system more complicated. To overcome this deficiency, the system is numerically solved by a MATLAB routine based on a numerical method bvp4c offered by Kierzenka and Shampine [22]. Many engineering problems has been successfully solved by this method.

## 4. Results and Discussion

The anisotropic slip, Soret, and Dufour effects on the flow, heat and mass transport properties of Williamson fluid driven by the infinite rotation of a rotating disk are studied by numerical solution of the governing equations. The investigation is extended out by determining the impact of pertinent parameters on fluid velocity, pressure, temperature, and concentration distribution presented through Figs. 2-23. The validity of the model and results is proved by comparison of the numerical values for a special case of William son fluid $(We = 0)$ in Tables 1 and 2. The numerical values of some important physical quantities are presented in Tables 3 and 4 for Williamson fluid. For simplicity, the constant values assumed for several parameters are $We = 0.4$, $\lambda_1 = 0.4$, $\lambda_2 = 0.2$, $M = 0.1$, $\text{Re} = 0.1$, $D_f = 0.5$, $S_r = 0.2$, $Sc = 1.0$, and $\text{Pr} = 6$.

The general structure of the radial, tangential, and axial velocity profiles observed from figures can be theoretically described as: the radial velocity profile shows the rapid growing behavior upward near the disk and then steadily diminishes to zero, allowing more fluid to pass through the disk, the tangential velocity profile appears to be exponentially decaying, and the axial velocity profile takes the asymptotic limiting value. Furthermore, the effects of the Weissenberg number on velocity profiles, and pressure distribution are presented in Figs. 2-5. Fig. 2 indicates that the Weissenberg number has increasing impact on the radial velocity near the disk that allows more fluid to pass through it, but it shows no effect as the velocity profile moves away from the disk. The tangential velocity shows a slightly decreasing behavior with an increase in $We$ as depicted in Fig. 3. While, the Weissenberg number is observed to have increasing influence on the axial velocity away from the disk as shown in Fig. 4. From Fig. 5, the pressure distribution is also observed to be increasing with an increase in $We$. The influence of radial slip on radial, tangential, and axial velocities in the presence of tangential slip is presented in Figs. 6-8 and the impact is quite similar to the effect of Weissenberg number. The radial slip increases the radial and axial velocities and slightly decreases the tangential velocity. While, the pressure, temperature, and concentration profiles decrease when the radial slip increases as shown in

Figs. 9-11. On the other hand, it is observed from Figs. 12-14 that the increase in tangential slip significantly reduces the radial and axial velocity profiles and shows a slight decreasing impact on tangential velocity near the disk. Figs. 15 and 16 depict the influence of magnetic field on radial and tangential velocities and both the profiles shows the decreasing behavior as the magnetic field gets stronger. The thermal-diffusion effects on heat and mass transfer are presented in Figs. 17 and 18. It can be clearly noted the Soret number reduces the heat transfer and increases the mass transfer. Figs. 19 and 20 depict the behavior of temperature and concentration profiles against Dufour number. The temperature distribution decreases near the disk and changes its pattern away from the disk whereas, the concentration profile slightly increases near the disk and reduces away from the disk. The Prandtl number decreases the heat transfer as indicated in Fig. 21. Finally, the influence of Schmidt number on temperature and concentration profiles is presented in Figs. 22 and 23 and it has the opposite impact on both the profiles, the temperature profile increases while the concentration profile decreases.

The Williamson fluid shows viscous behavior when $We = 0$ as $\mu_0 \rightarrow \mu_\infty$ or $\Gamma = 0$. Considering this property of Williamson fluid, the model and the results are verified by comparing them with the results obtained in [23] for viscous flows over a rotating disk as presented in Table 1. Moreover, in Table 2, the numerical values of heat transfer are compared for different values of the Prandtl number with the results obtained in [24]. The numerical comparison in both the tables proves the validity of the results. The magnitude of the numerical values of radial and tangential skin-friction coefficients and coefficient of friction are presented in Table 3 whereas, the magnitude of the numerical values of local Nusselt number and the local Sherwood number is given in Table 4. From table 3, it is observed that the Weissenberg number has the similar impact on the magnitude of skin-friction drag in both radial and tangential directions and moment of friction, all the quantities increase as $We$ increases. The effects of radial slip on radial and tangential skin-friction drags are opposite to the tangential slip effects. The radial slip increases the skin-friction in both directions increasing the moment of friction while the tangential slip decreases both the skin-friction coefficients and the moment of friction. However, the skin-friction in radial direction reduces as the magnetic field gets stronger, but enhances the drag in tangential direction, increasing the moment of friction. Moreover, it is observed from Table 4 that the increase in $We$ increases both the heat and mass transport in the fluid, the radial slip also increases the local Nusselt and Sherwood numbers, but the tangential slip and the magnetic field parameters decrease both the quantities. On the other hand, the effects of Prandtl, Dufour, and Soret numbers are different on Nusselt and Sherwood numbers. The Prandtl and Soret numbers increase the heat transfer and reduce the mass transport, but the Dufour number decreases the heat transfer and increases the mass transport in flows of Williamson fluid over a rotating disk in the presence of anisotropic slip.

## 5. Conclusion

In this paper, an effort is made towards the numerical investigation of Soret and Dufour effects on MHD flow of Williamson fluid over an infinite rotating disk with anisotropic slip. The main focus of the investigation was on the anisotropic slip and the Soret and Dufour effects which greatly influence the flow, heat and mass transfer properties. The PDEs governing these characteristics were transformed into ODEs by using von Kármán's similarity transformation. The resulting ODEs were highly nonlinear and the analytical solution was not possible because of the presence of slip condition. Therefore, the results were obtained by using a numerical scheme bvp4c and presented through graphs and tables. The radial slip increased the radial velocity and the tangential slip decreased the velocity in this direction. The magnetic flield decresed both the radial and azimuthal velocities. The efficiency of the numerical results was also confirmed by comapring them with the published results found in literature. Moreover, the numerical values of some physical quantities were also obtained and presneted through tables.

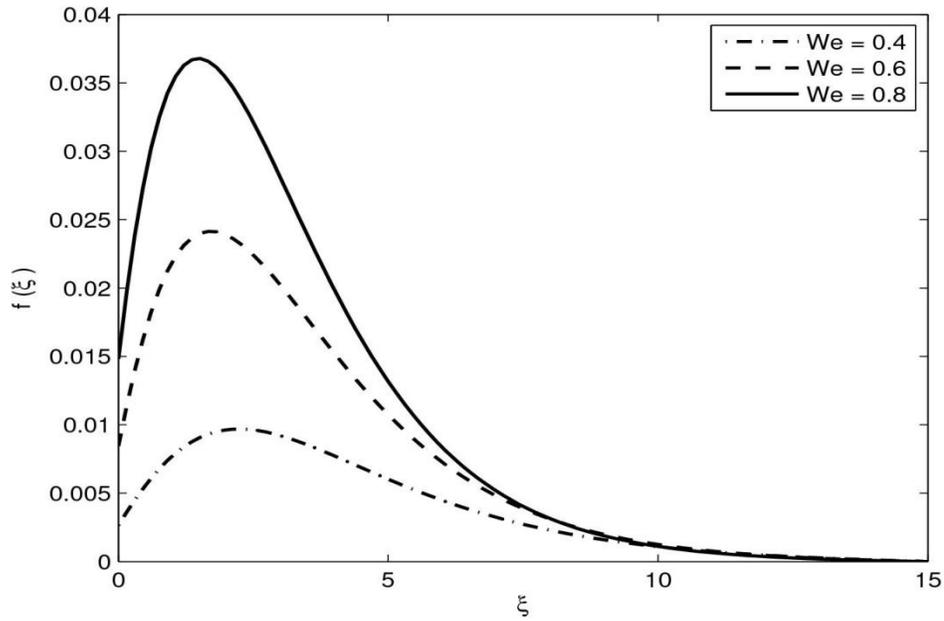

Fig. 2: The influence of Weissenberg number on radial velocity in the presence of anisotropic slip

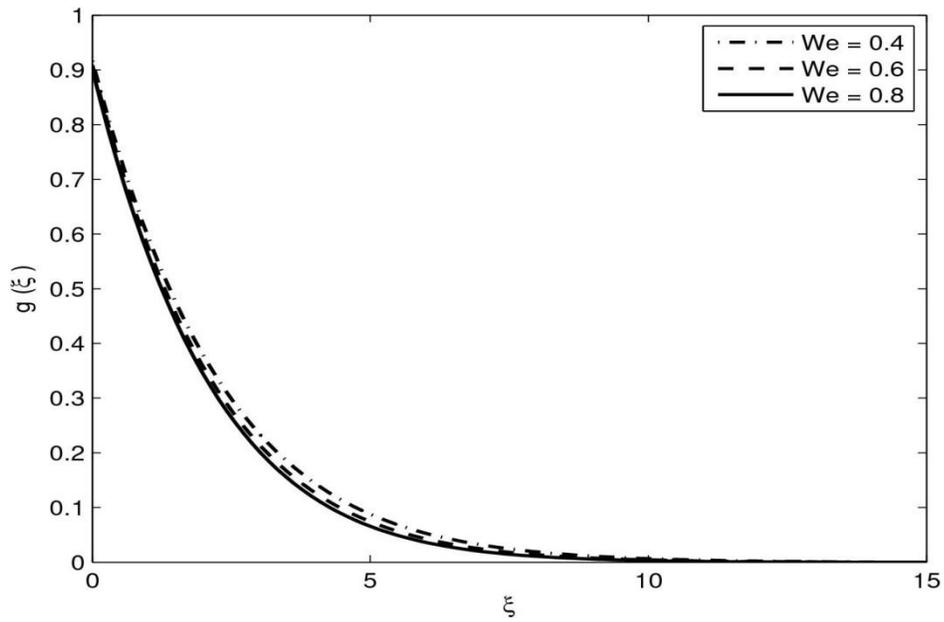

Fig. 3: The influence of Weissenberg number on tangential velocity in the presence of anisotropic slip

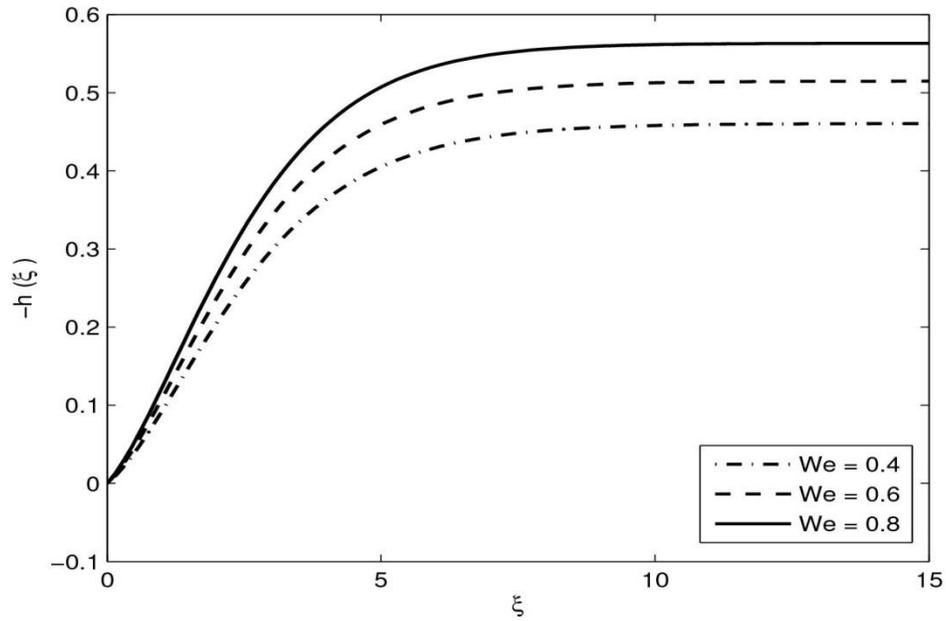

Fig. 4: The influence of Weissenberg number on axial velocity in the presence of anisotropic slip

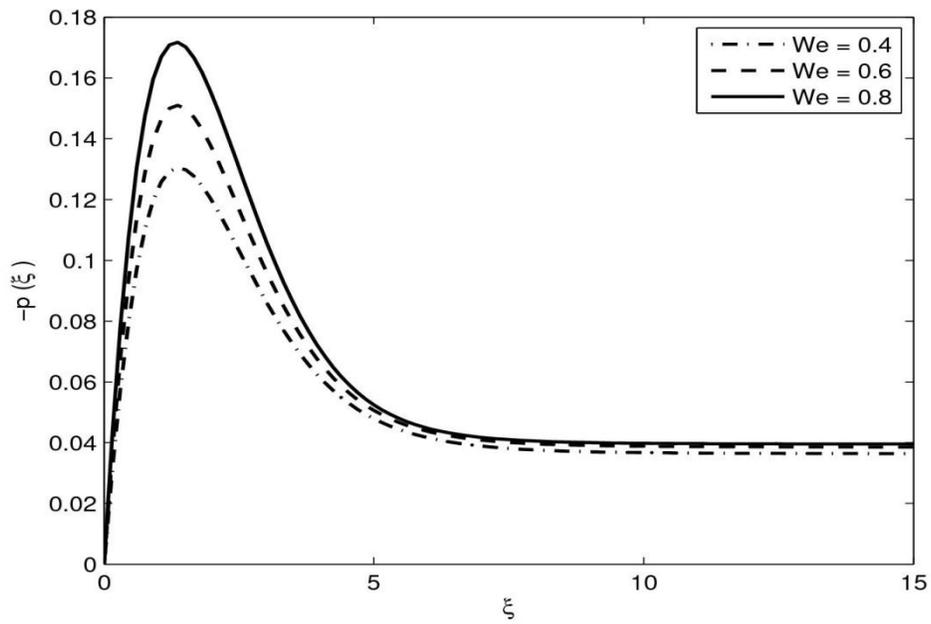

Fig. 5: The influence of Weissenberg number on axial velocity in the presence of anisotropic slip

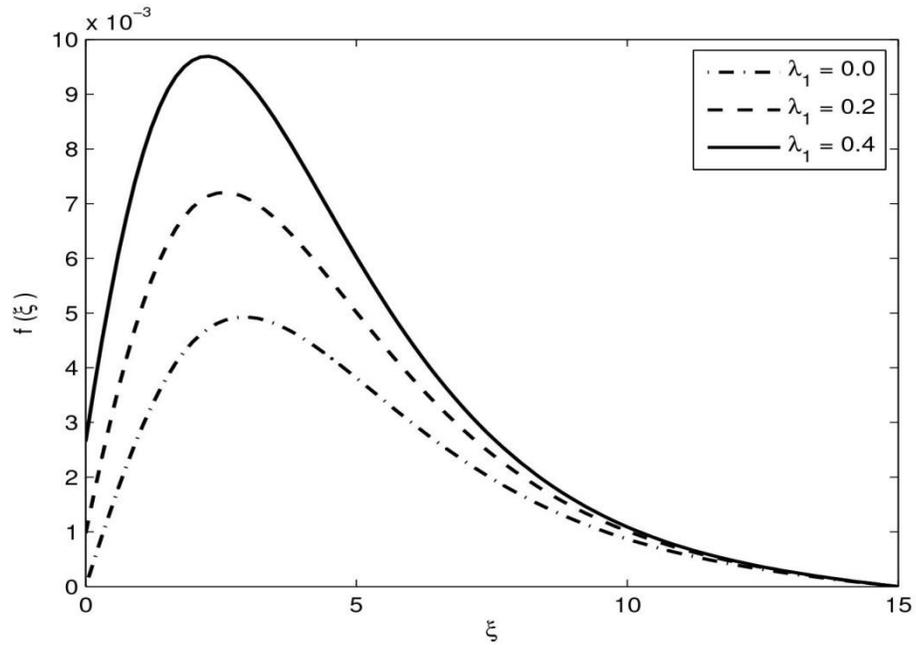

Fig. 6: The influence of radial slip on radial velocity in the presence of tangential slip

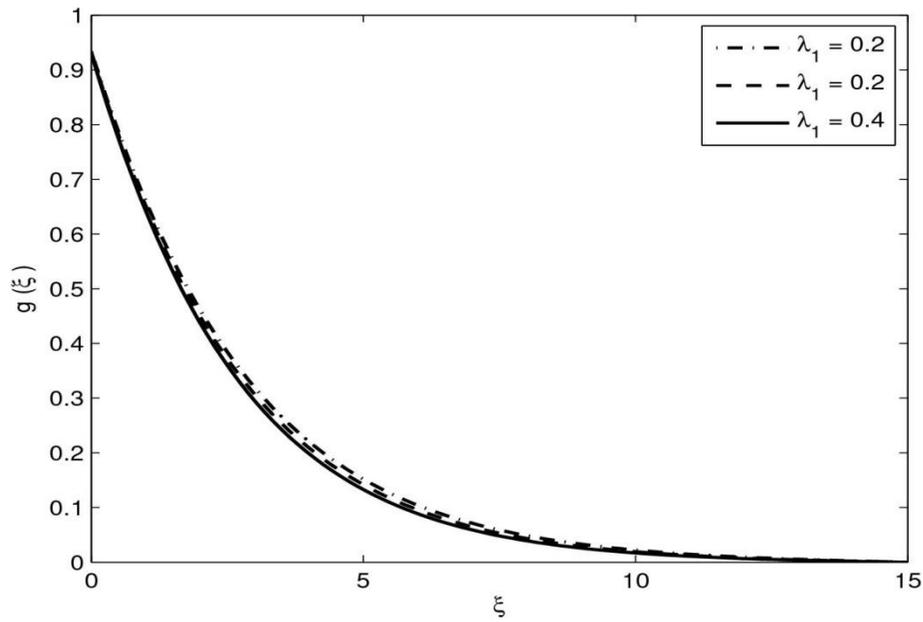

Fig. 7: The influence of radial slip on tangential velocity in the presence of tangential slip

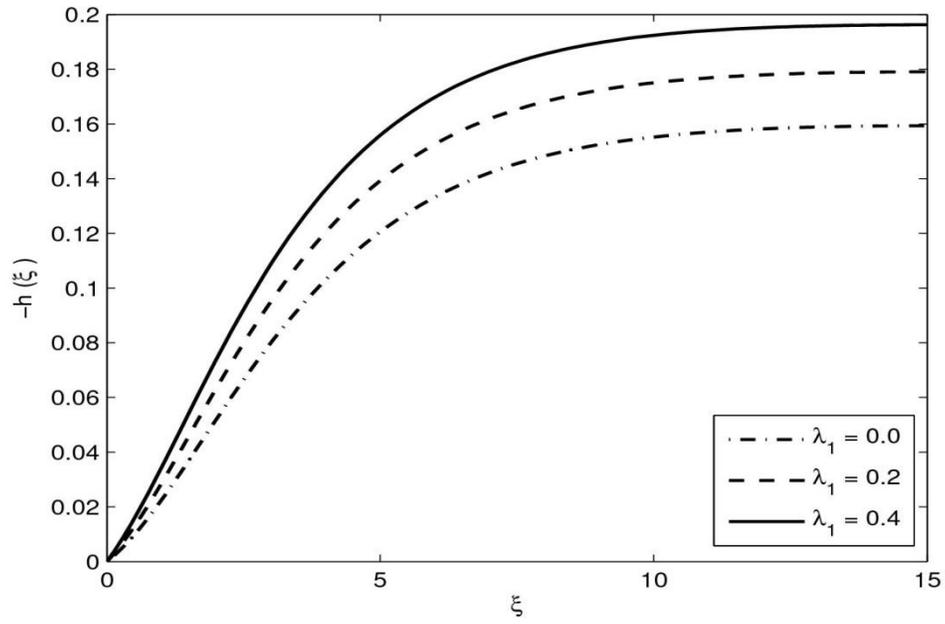

Fig. 8: The influence of radial slip on axial velocity in the presence of tangential slip

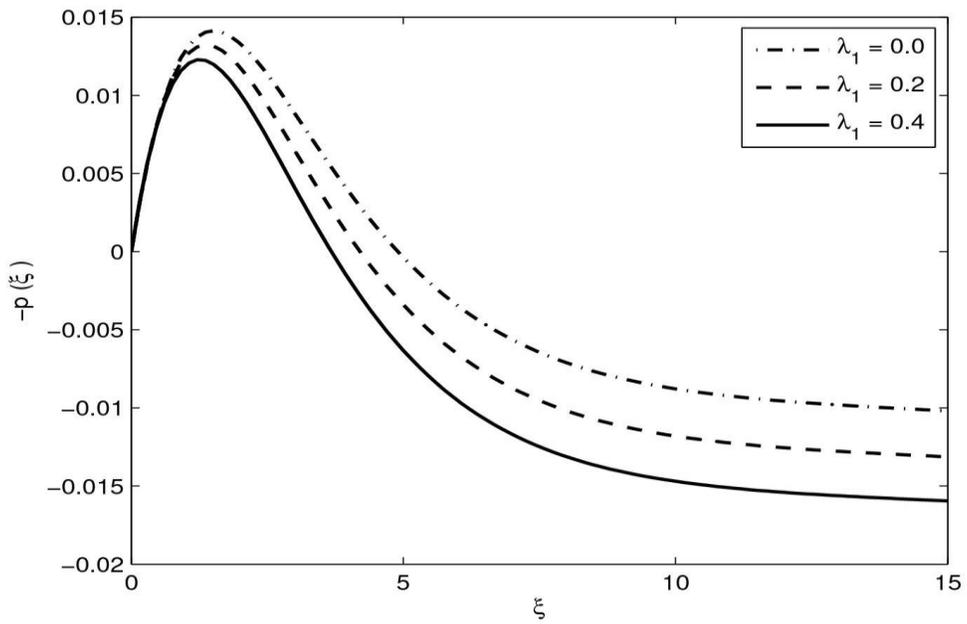

Fig. 9: The influence of radial slip on pressure distribution in the presence of tangential slip

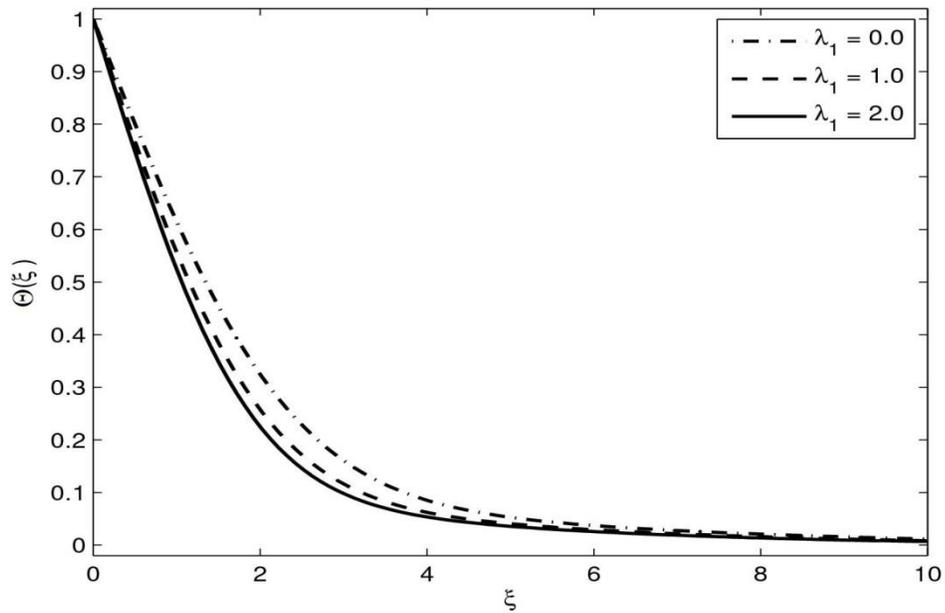

Fig. 10: The influence of radial slip on temperature distribution in the presence of tangential slip

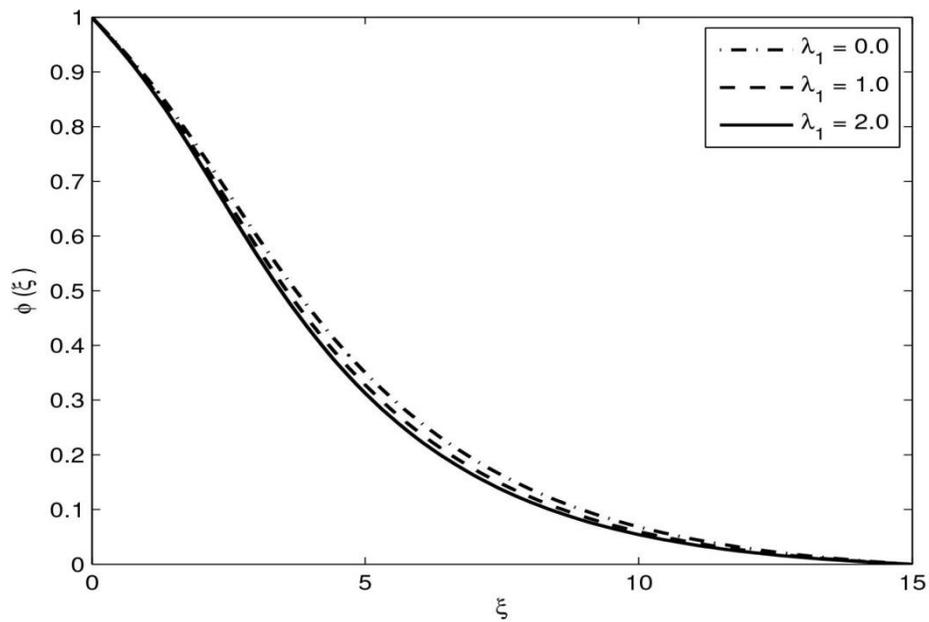

Fig. 11: The influence of radial slip on concentration distribution in the presence of tangential slip

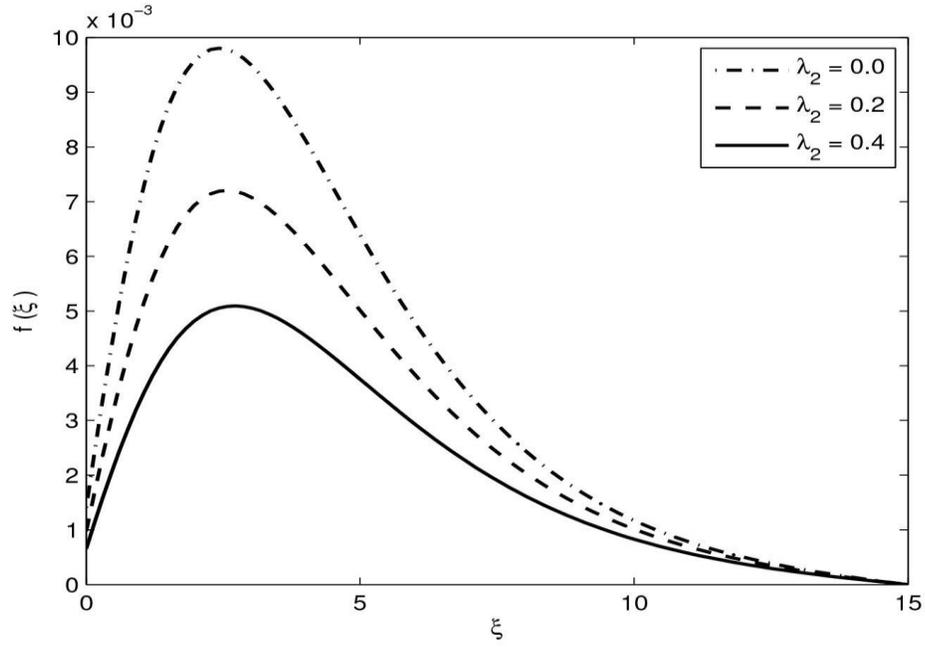

Fig. 12: The influence of tangential slip on radial velocity profile in the presence of radial slip

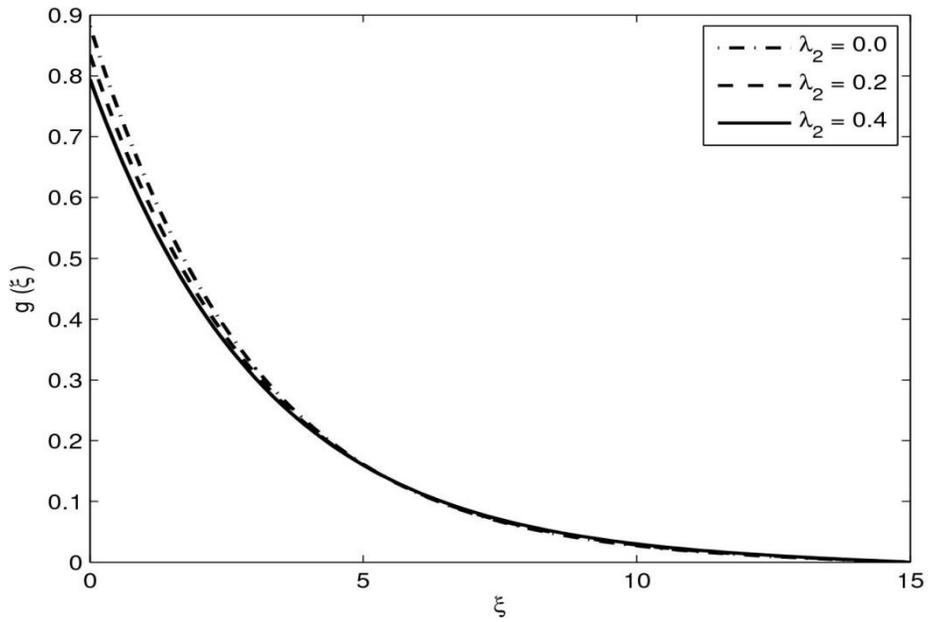

Fig. 13: The influence of tangential slip on tangential velocity profile in the presence of radial slip

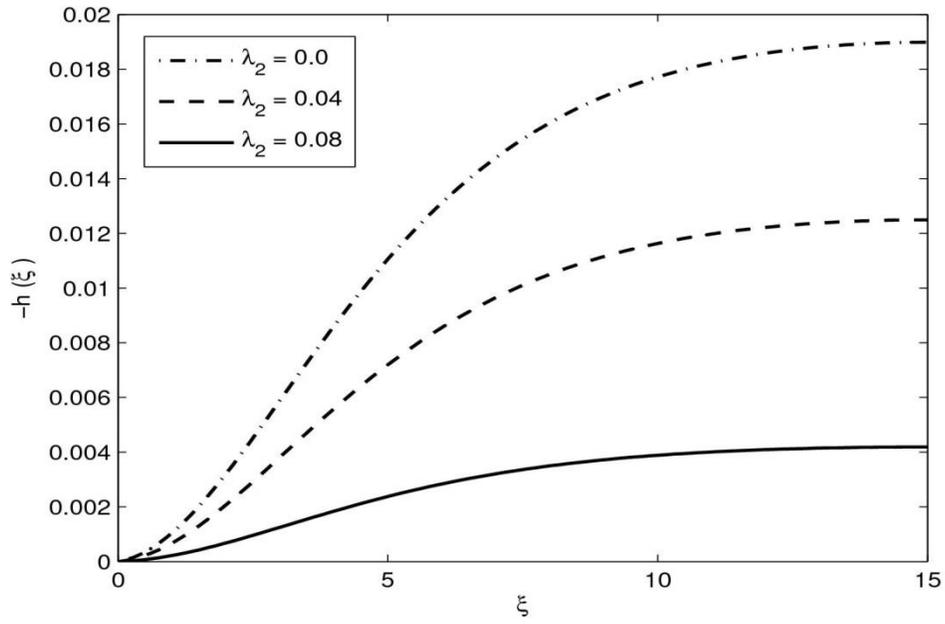

Fig. 14: The influence of tangential slip on axial velocity profile in the presence of radial slip

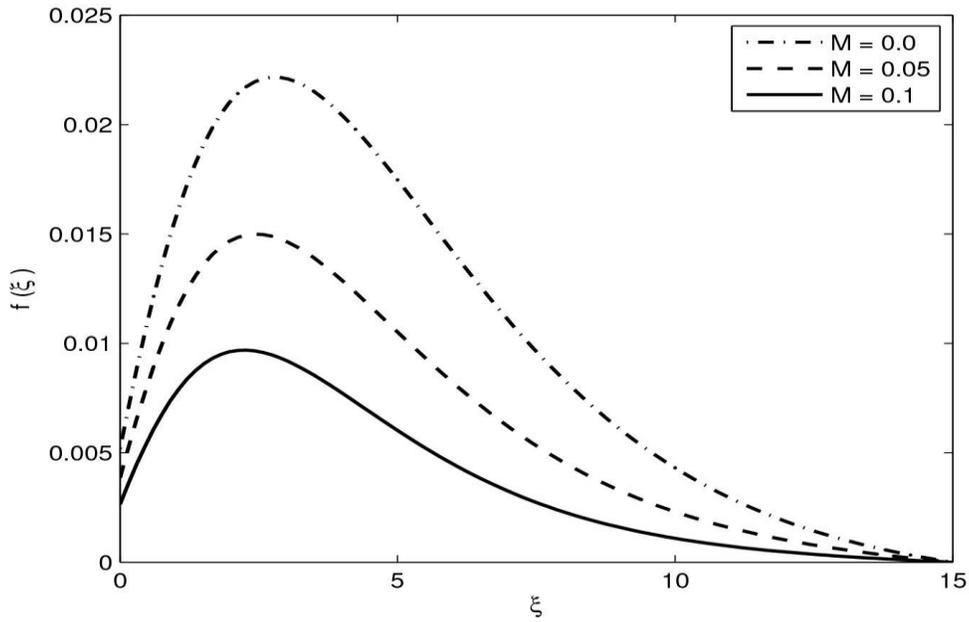

Fig. 15: The influence of magnetic field on radial velocity profile in the presence of anisotropic slip

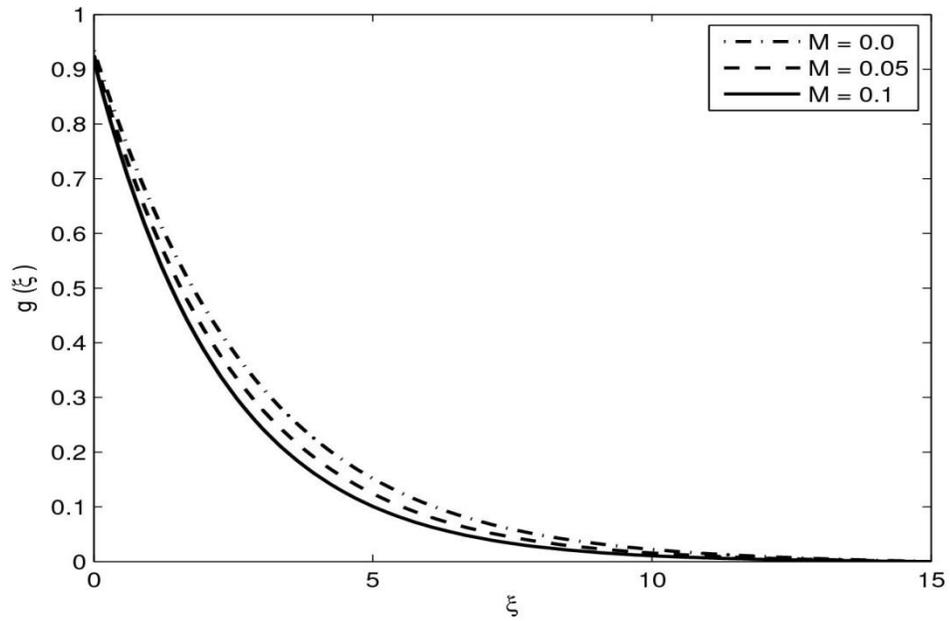

Fig. 16: The influence of magnetic field on tangential velocity profile in the presence of anisotropic slip

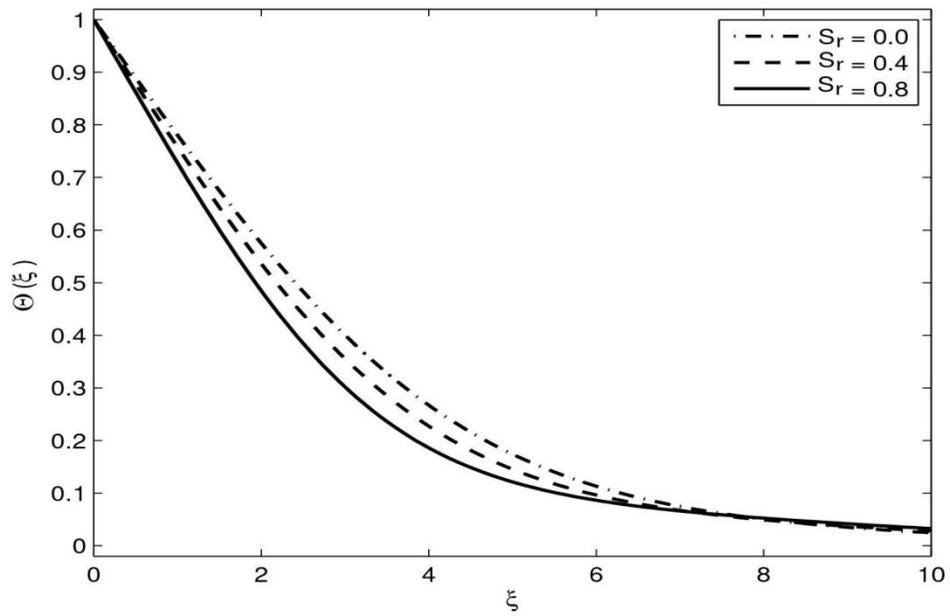

Fig. 17: The influence of Soret parameter on temperature distribution in the presence of anisotropic slip

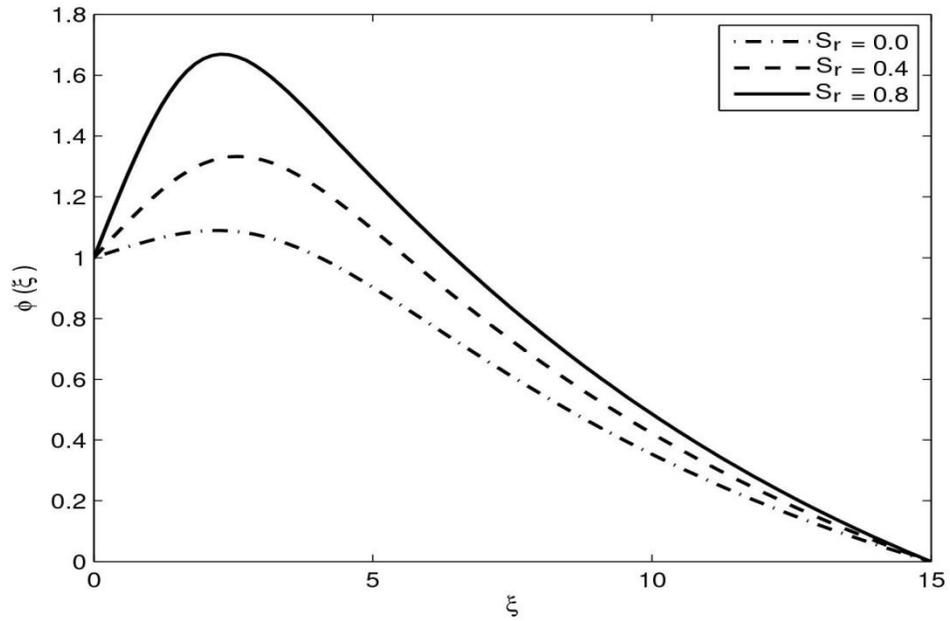

Fig. 18: The influence of Soret parameter on concentration distribution in the presence of anisotropic slip

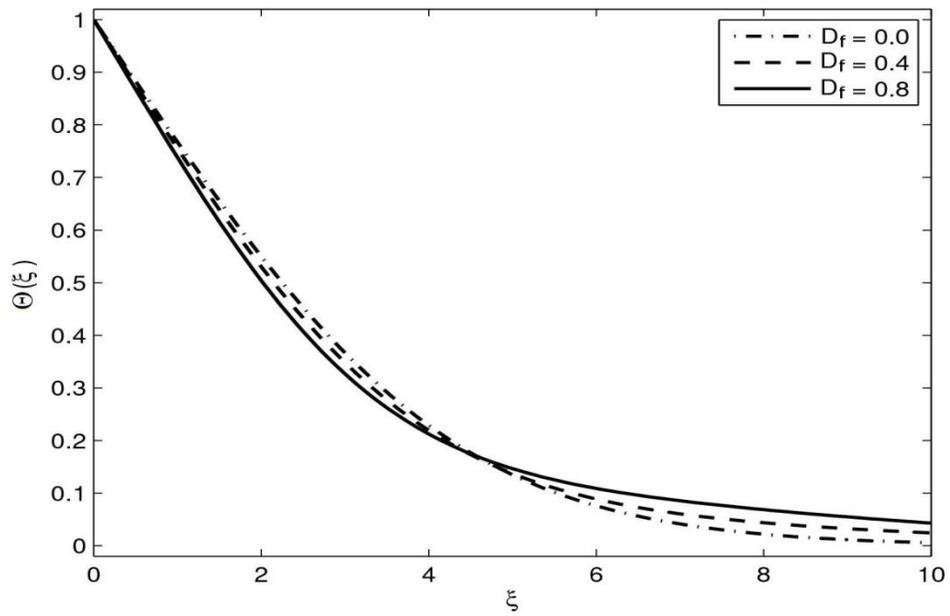

Fig. 19: The influence of Dufour parameter on temperature distribution in the presence of anisotropic slip

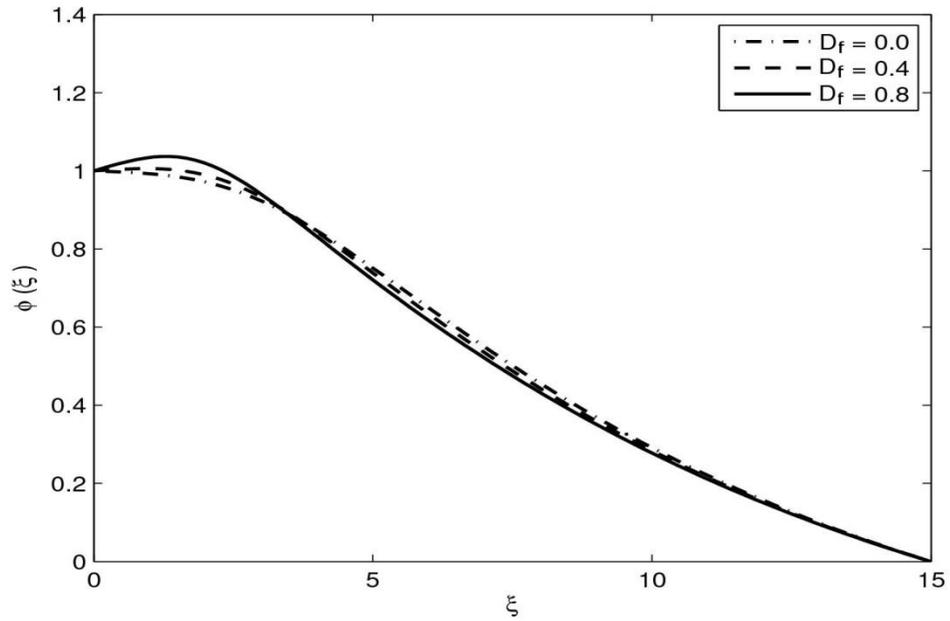

Fig. 20: The influence of Dufour parameter on concentration distribution in the presence of anisotropic slip

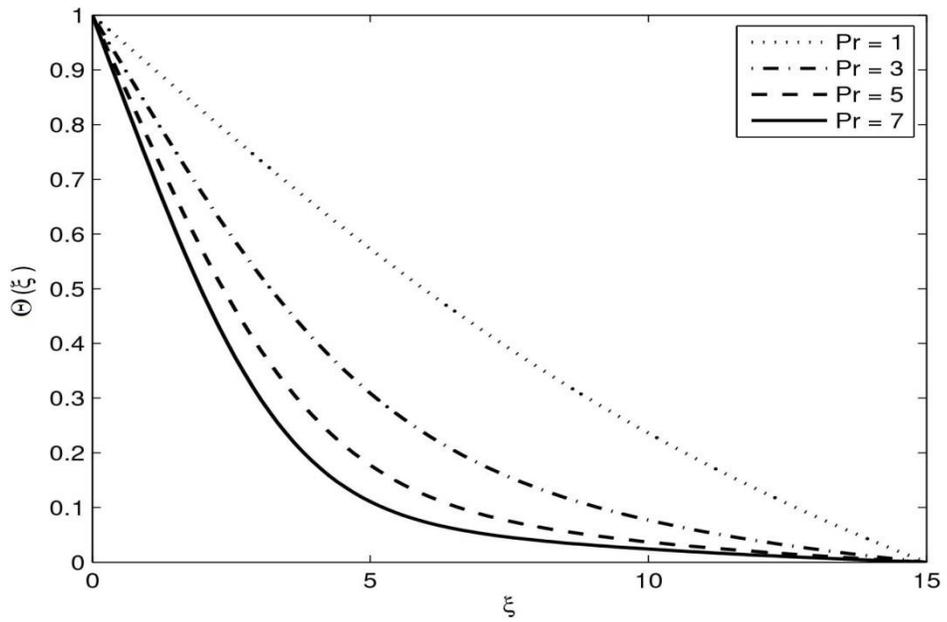

Fig. 21: The influence of Prandtl number on temperature distribution in the presence of anisotropic slip

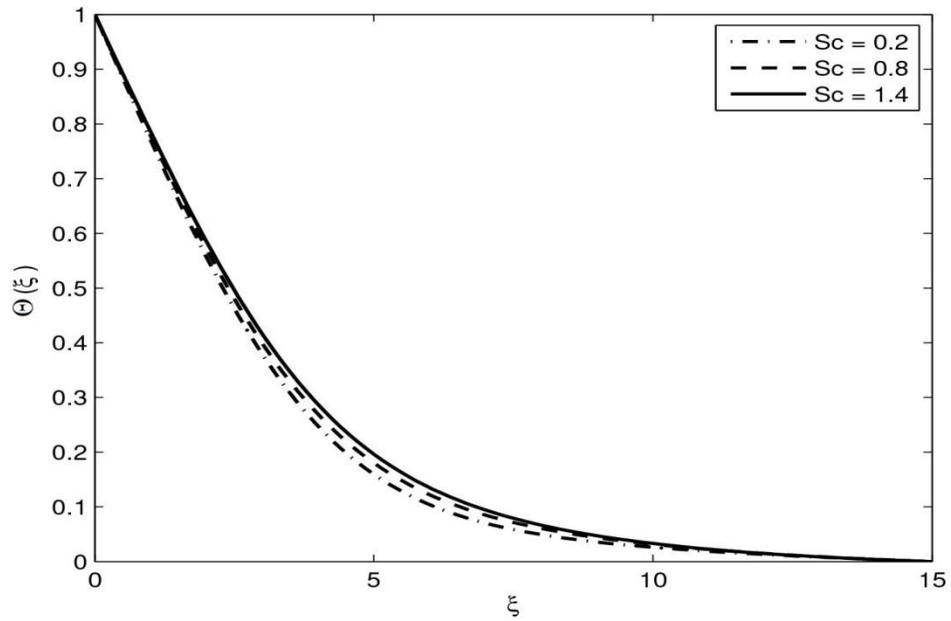

Fig. 22: The influence of Schmidt number on temperature distribution in the presence of anisotropic slip

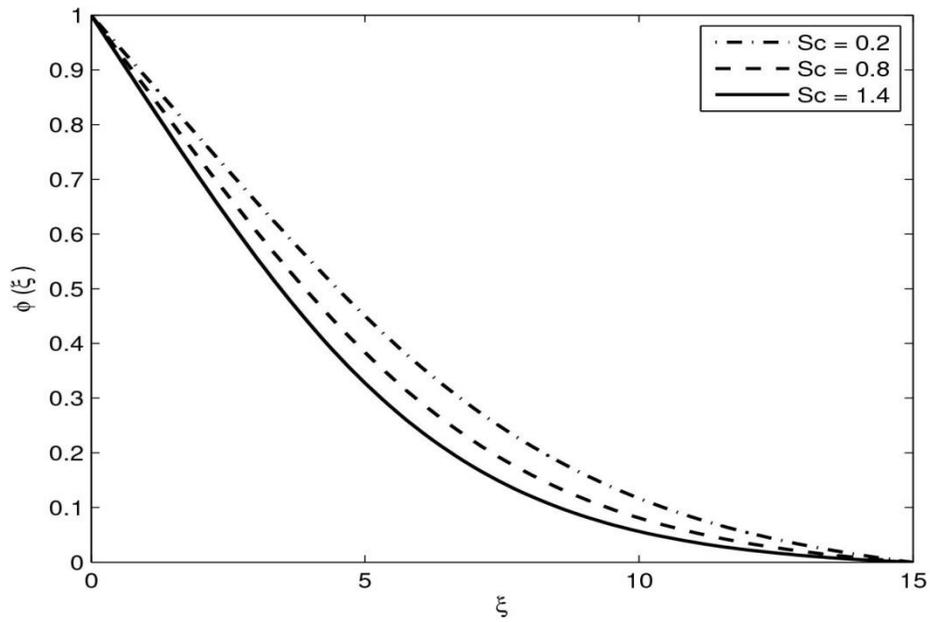

Fig. 23 The influence of Schmidt number on concentration distribution in the presence of anisotropic slip

Table-1: Numerical comparison of some viscous physical quantities with Ref. [23] obtained at $We = 0$, $M = 0$, $\Pr = 0.71$, $Sc = 0.6$, $D_f = 0$, $S_r = 0$

| Physical Quantities | Present | [23] |
|---|---|---|
| $f'(0)$ | 0.5102 | 0.51023 |
| $-g'(0)$ | 0.6159 | 0.61592 |
| $-h(\infty)$ | 0.8844 | 0.8838 |
| $-P(\infty)$ | 0.3911 | 0.3906 |

Table-21: Numerical comparison of $-\Theta'(0)$ with Ref. [24] obtained at $We = 0$, $M = 0$, $Sc = 0.6$, $D_f = 0$, $S_r = 0$

| Pr | $-\Theta'(0)$ | [24] |
|---|---|---|
| 0.71 | 0.3286 | 0.32857 |
| 1 | 0.3962 | 0.39626 |
| 10 | 1.1341 | 1.1341 |
| 100 | 2.6871 | 2.8672 |

Table 3: Numerical values of local skin friction coefficients $C_f$ and $C_g$, and moment of friction $C_m$ for $We$, $\lambda_1$, $\lambda_2$, and $M$ while keeping $\mathrm{Re} = 0.1$, $D_f = 0.5$, $S_r = 0.2$, $Sc = 1.0$, and $\Pr = 6$

| $We$ | $\lambda_1$ | $\lambda_2$ | $M$ | $\mathrm{Re}^{-\frac{1}{2}} C_f$ | $\mathrm{Re}^{-\frac{1}{2}} C_g$ | $C_m$ |
|---|---|---|---|---|---|---|
| 0.4 | | | | 0.0066 | 0.3295 | 6.5461 |
| 0.6 | 0.4 | 0.2 | 0.1 | 0.0211 | 0.3758 | 7.4677 |
| 0.8 | | | | 0.0371 | 0.4176 | 8.2980 |
| | 0.0 | | | 0.0032 | 0.3126 | 6.2110 |
| 0.4 | 0.2 | 0.2 | 0.1 | 0.0049 | 0.3203 | 6.3633 |
| | 0.4 | | | 0.0066 | 0.3295 | 6.5461 |
| | | 0.0 | | 0.0092 | 0.3636 | 7.2253 |
| 0.4 | 0.4 | 0.2 | 0.1 | 0.0066 | 0.3295 | 6.5461 |
| | | 0.4 | | 0.0047 | 0.3025 | 6.0105 |

| | | | 0.0 | 0.0130 | 0.2166 | 4.3032 |
|---|---|---|---|---|---|---|
| 10 | 0.2 | 0.2 | 0.05 | 0.0097 | 0.2758 | 5.4791 |
| | | | 0.1 | 0.0066 | 0.3295 | 6.5461 |

Table 4: Numerical values of local Nusselt number $N_u$ and local Sherwood number $S_h$ for physical parameters $We, \lambda_1, \lambda_2, M, \text{Pr}, D_f,$ and $S_r$ while keeping $\text{Re} = 0.1$

| $We$ | $\lambda_1$ | $\lambda_2$ | $M$ | Pr | $D_f$ | $S_r$ | $\text{Re}^{-\frac{1}{2}} N_u$ | $\text{Re}^{-\frac{1}{2}} S_h$ |
|---|---|---|---|---|---|---|---|---|
| 0.4 | | | | | | | 0.2319 | 0.0686 |
| 0.6 | 0.4 | 0.2 | 0.1 | 6 | 0.5 | 0.2 | 0.3692 | 0.0924 |
| 0.8 | | | | | | | 0.4603 | 0.1148 |
| | 0.0 | | | | | | 0.1571 | 0.0651 |
| 0.4 | 0.2 | 0.2 | 0.1 | 6 | 0.5 | 0.2 | 0.1944 | 0.0663 |
| | 0.4 | | | | | | 0.2319 | 0.0686 |
| | | 0.0 | | | | | 0.2664 | 0.0732 |
| 0.4 | 0.4 | 0.2 | 0.1 | 6 | 0.5 | 0.2 | 0.2319 | 0.0686 |
| | | 0.4 | | | | | 0.2009 | 0.0659 |
| | | | 0.0 | | | | 0.3291 | 0.0988 |
| 0.4 | 0.4 | 0.2 | 0.05 | 6 | 0.5 | 0.2 | 0.2823 | 0.0809 |
| | | | 0.1 | | | | 0.2319 | 0.0686 |
| | | | | 1 | | | 0.0851 | 0.0965 |
| 0.4 | 0.4 | 0.2 | 0.1 | 3 | 0.5 | 0.2 | 0.1555 | 0.0829 |
| | | | | 5 | | | 0.2096 | 0.0728 |
| | | | | | 0.0 | | 0.2346 | 0.0681 |
| | | | 0.1 | | 0.5 | 0.2 | 0.2319 | 0.0686 |
| | | | | | 1.0 | | 0.2288 | 0.0693 |
| | | | | | | 0.0 | 0.2212 | 0.1003 |
| 0.4 | 0.4 | 0.2 | 0.1 | 6 | 0.5 | 0.5 | 0.2319 | 0.0686 |
| | | | | | | 1.0 | 0.2442 | 0.0324 |